\documentclass[%
 aip,
 amsmath,amssymb,
 reprint,%
]{revtex4-1}

\usepackage{graphicx}
\usepackage{dcolumn}
\usepackage{bm}

\usepackage[utf8]{inputenc}
\usepackage[T1]{fontenc}
\usepackage{mathptmx}

\begin{document}

\preprint{AIP/123-QED}

\title[]{High-performance transmission of optical vortices through scattering media with opaque-lens field transformation theory}

\author{Hengkang Zhang}
 
\author{Bin Zhang}%
 \email{zhangbin1931@126.com.}
\author{Zhensong Wan}
 
\author{Qiang Liu}
 \email{qiangliu@mail.tsinghua.edu.cn.}
\affiliation{ 
Key Laboratory of Photonics Control Technology (Tsinghua University), Ministry of Education, Beijing 100084, China
}%
\affiliation{ 
State Key Laboratory of Precision Measurement Technology and Instruments, Department of Precision Instrument, Tsinghua University, Beijing 100084, China
}%

\date{\today}

\begin{abstract}
The applications of vortex beam in biomedical field are always hindered by light scattering of biological tissues. Recently wavefront shaping technology has shown the potential in focusing a structured light field through scattering media. Herein we have proposed a novel theory to describe the field transformation property of the opaque lens. This theory makes it possible to generate foci with specified amplitude and phase profiles by well-designing the incident field. We verify the validity of this theory by experimentally generating doughnut-shaped foci with various topological charges, and the foci match well with the theoretical predictions. Meanwhile, we have also analyzed the limitations of our approach, and it reveals that there is a trade-off between the thickness of the medium and the acceptance spatial bandwidth of the incident field. Our approach will inspire new research ideas for the deep-tissue applications of vortex beam in optics and dynamics, which is significant to numerous potential bio-photonics technologies.
\end{abstract}

\maketitle

%

\section{\label{sec1}Introduction}

Vortex beam is characterized by the helical phase structure, which is equivalent to carry orbital angular momentum (OAM) in the quantum theory. The excellent optical and mechanical properties of vortex beam make it become one of the most attractive beams in numerous areas \cite{Shen2019,Forbes2019,Yao2011161}. The application of vortex beam in biomedical optics has enabled the development of a series of novel and practical technologies, such as multiple-degree-of-freedom optical tweezers\cite{Jeffries2007415}, super-resolution microscopy\cite{Willig2006935,Kozawa201886}, single-cell nanosurgery\cite{Asavei2013}, and so on. However, it is still a grand challenge to transmit vortex beam deep through biological tissues, as the inhomogeneous refractive index strongly scatters the vortex beam and disrupts the phase structure\cite{Shi20171756,Wang20162069}. The intricate interference of the scattered light results in the formation of a random light field, which is well known as speckle pattern.

With the further understanding of the scattering mechanism, it is revealed that light scattering in a linear and static medium is not totally disordered, but a deterministic process. On the basis of this nature, wavefront shaping technology with spatial light modulator (SLM) has developed into an effective way for manipulating light propagation inside scattering media, such as focusing light to desired target locations\cite{Vellekoop20072309,Vellekoop201512189,Lai2015126,Tzang20176122,Luo2020} or coupling light into open channels\cite{Choi2011,Jo2017,YH352}. Nowadays, the scattering medium is gradually accepted as a class of optical element for light focusing. It is also called “opaque lens”\cite{Rotter2017}. 

There are two important foundations for the investigations and applications of opaque lens. One is the “memory effect” of scattering media, which describes the tilt/shift correlations of the transmission light\cite{Freund19882328,Judkewitz2015684,Osnabrugge2017886}. Once the optimized wavefront for deep focusing is obtained, the incident light that is tilted inside an angular range can also be focused, resulting in a continuous lateral scanning of the focus. This angular range determined by the memory effect serves as the field of view of the opaque lens. The other is the shape and size of the focus generated by the opaque lens. In light of the equation derived by I. M. Vellekoop et al., the focused field is proportional to the Fourier transformation of the intensity profile on the back surface of the medium\cite{Vellekoop2010320}. This equation gives a theoretical prediction of the diffraction limitation of the opaque lens. In the past decade, opaque lens has shown its practicability in single-spot and multiple-spot focusing\cite{Vellekoop20072309,Zhang2018,Popoff2010}, as well as wide-field imaging\cite{Katz2012549,Li20192845}. However, most of these demonstrations aimed to focus a flat-top beam or image an intensity pattern, and the phase information was usually not concerned. There still lacks a theoretical framework describing the response of the opaque lens to a field with a certain phase structure, although it is crucial to the purpose of focusing vortex beam through scattering media.

Recently, a remarkable research was reported by A. Boniface et. al.\cite{Boniface201754}. They generated several different types of foci, including the Bessel, “donut”, and double helix foci, by designing the point spread function (PSF) of the opaque lens. The PSF was controlled by applying a well-chosen mask in the virtual Fourier plane of the output modes.

In this article, we explore the generation of the foci with specified amplitude and phase structures based on the field transformation property of opaque lens. We begin with the derivation of a theoretical framework, which predicts the spatial profile of the focus for any an interested scalar incident field. Then, to verify the validity of the theory, we conducted the experiments of single-spot and multiple-spot focusing of Laguerre-Gaussian beams with various topological charges and achieved doughnut-shaped foci behind the scattering medium. The topological charges of Laguerre-Gaussian beams were identified by the optical-lattice diffraction patterns created with a triangle aperture. We believe that this theory provides a description of the generalized focusing property of opaque lens, and our technology will be significant to the applications of vortex beam in scattering environment.

\section{\label{sec2}Theory}

\subsection{\label{sec1_1}Opaque-lens field transformation theory}

Considering a simplified configuration of light focusing by opaque lens. As shown in Fig. 1(a), the linearly polarized light propagates through a scattering medium, then the transmission light is measured in an observation plane. The medium is perpendicular to the optical axis, and a modulation plane is placed at the front surface of the medium. Without optimization, the incident wavefront is disrupted by light scattering, resulting in a random speckle pattern at the observation plane. The auto-correlation function of the speckle pattern is written as\cite{Goodman07}
\begin{equation}
\Gamma ^{obs}\left( {x,y,\Delta x,\Delta y} \right) = \left\langle {{E^{obs}}\left( {x,y} \right){E^{obs * }}\left( {x - \Delta x,y - \Delta y} \right)} \right\rangle,
\end{equation}
where $E^{obs}$ represents the field (amplitude and phase) on the observation plane, and the angle bracket donates the ensemble average. Then we consider the field $E^{b}$ at the back surface of the medium. For practical purposes with a medium dominated by multiple-scattering in the diffused region, a general approximation of $E^{b}$ is that its correlation extent is sufficiently small. Then the auto-correlation function of $E^{b}$ could be expressed as a 2-D delta function.
\begin{equation}
\Gamma ^b\left( {\xi ,\eta ,\Delta \xi ,\Delta \eta } \right) = \kappa {I^b}\left( {\xi ,\eta } \right)\delta \left( {\Delta \xi ,\Delta \eta } \right),
\end{equation}
where $\kappa$ is a constant coefficient, and ${I^b}\left(\xi,\eta\right)={E^b}\left(\xi,\eta\right){E^{b*}}\left(\xi,\eta\right)$ donates the intensity at the point $\left(\xi,\eta\right)$ on the back surface of the medium. With this approximation, the relation between the auto-correlation function $\Gamma^{obs}$ and the intensity profile $I^{b}$ is described by Van Cittert–Zernike theorem\cite{Goodman15}. This theorem reveals that $\Gamma^{obs}$ is the 2-D Fourier transformation of the intensity profile $I^{b}$,
\begin{eqnarray}
{\Gamma ^{obs}}&&\left( {\Delta x,\Delta y} \right) = \frac{\kappa }{{{\lambda ^2}{z^2}}} \nonumber\\
&&\times 
\iint_{S^b}{I^b}\exp \left[ { - \frac{{i2\pi }}{{\lambda z}}\left( {\xi \Delta x + \eta \Delta y} \right)} \right]{\rm{d}}\xi {\rm{d}}\eta,
\end{eqnarray}
where $\lambda$ donates the wavelength, $S^{b}$ donates the area over which the transmission light distributes, and \emph{z} donates the distance between the observation plane and the back surface of the medium.

\begin{figure}
\includegraphics[width=8.5cm]{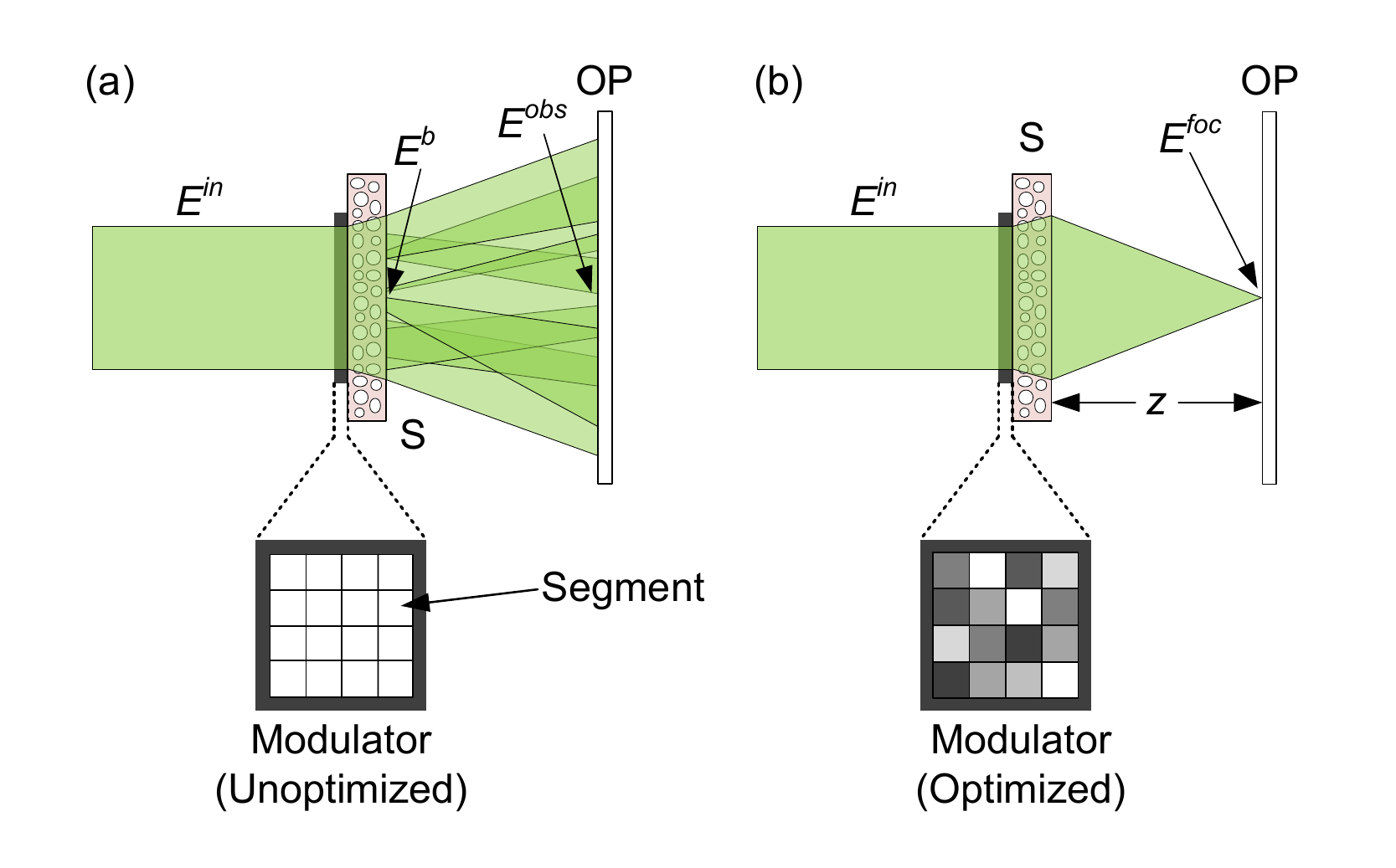}
\caption{\label{fig:1} A simplified configuration of light focusing by opaque lens. (a) Without optimization, the incident light is scattered and forms a speckle pattern. (b) After the wavefront is optimized, the incident light is focused. S, scattering medium; OP, observation plane.}
\end{figure}

To perform a digital control over the incident wavefront, the modulator is divided into \emph{N} identical segments. With the Fresnel-Kirchhoff diffraction formula, the transmission coefficient that relates the \emph{n}-th segment and an arbitrary point $\left( {x,y} \right)$ on the observation plane is calculated as
\begin{equation}
{t_n}\left( {x,y} \right) = \frac{i}{\lambda }
\iint_{S_n^{in}}
{g\left( {\xi ,\eta ,x,y} \right){\rm{d}}\xi {\rm{d}}\eta },
\end{equation}
where $S_n^{in}$ represents the area of a single segment, and $g\left( {\xi ,\eta ,x,y} \right)$ represents the Green function of the propagation from the front surface of the medium to the observation plane. Assuming that the modulator could provide amplitude-phase modulation to the incident beam, then a focus is generated at the point $\left( {x,y} \right)$ by multiplying the field at each of the segments by a factor of $t_n^*\left({x,y}\right)$, as shown in Fig. 1(b). Tilting the incident beam by an angle will result in the correlated transverse shift of the focus, and the acceptable angular range of tilting is determined by the memory effect of the scattering medium. Since each tilting angle corresponds to a spatial spectral component, the acceptance angle of the memory effect poses a fundamental limitation to the spatial frequency spectrum of the incident field, and the spatial spectral component that exceeds the acceptance angle will form speckles rather than a focus. As a consequence of the finite spatial frequency, the amplitude and phase of the incident field vary slowly in the spatial domain, and an approximation of the incident field is available when the area of a segment is sufficiently small. That is, for an arbitrary function $f\left( {\xi ,\eta} \right)$, the incident field $E^{in}\left( {\xi ,\eta} \right)$ satisfies
\begin{eqnarray}
\iint_{S_n^{in}}&&
{{E^{in}}\left( {\xi ,\eta } \right)f\left( {\xi ,\eta } \right){\rm{d}}\xi {\rm{d}}\eta } \nonumber\\
&&
\approx E^{in}\left(\xi _n^c,\eta _n^c\right) \iint_{S_n^{in}}{f\left( {\xi ,\eta } \right){\rm{d}}\xi {\rm{d}}\eta },
\end{eqnarray}
where $\left(\xi _n^c,\eta _n^c\right)$ donates the center coordinates of the \emph{n}-th segment. We can now calculate the focused field for an incident beam with interested amplitude and phase profiles. Without loss of generality, the point $\left(0,0\right)$ is chosen as the focal point, and the focused field in the vicinity of the point $\left(0,0\right)$ is expressed as
\begin{equation}
{E^{foc}}\left( {x,y} \right) = \sum\limits_{n = 1}^N {\frac{i}{\lambda }\iint_{S_n^{in}}{E^{in}}t_n^ * \left( {0,0} \right)g\left( {\xi ,\eta ,x,y} \right){\rm{d}}\xi {\rm{d}}\eta } .
\end{equation}

By applying the approximation of Eq. (5) to Eq. (6), and taking the ensemble average of the focused field, we rewrite Eq. (6) as
\begin{eqnarray}
&& \left\langle {E^{foc}}\left( {x,y} \right) \right\rangle \nonumber\\
&&  =\left\langle {\sum\limits_{n = 1}^N {{E^{in}}\left( {\xi _n^c,\eta _n^c} \right)t_n^ * \left( {0,0} \right)\frac{i}{\lambda }\iint_{S_n^{in}} g\left( {\xi ,\eta ,x,y} \right){\rm{d}}\xi {\rm{d}}\eta } } \right\rangle \nonumber\\
&&  =\sum\limits_{n = 1}^N {{E^{in}}\left( {\xi _n^c,\eta _n^c} \right)\left\langle {t_n^ * \left( {0,0} \right){t_n}\left( {x,y} \right)} \right\rangle } .
\end{eqnarray}
Another interpretation of the transmission coefficient ${t_n}\left({x,y}\right)$ is the response of $E_n^{seg}$ on the observation plane, where $E_n^{seg}$ is defined as an unit-amplitude uniform field which has the same cross section as the \emph{n}-th segment. Therefore, the term $\left\langle {t_n^ * \left( {0,0} \right){t_n}\left( {x,y} \right)} \right\rangle$ is the auto-correlation function which takes the form of Eq. (3). With Eqs. (3) and (7) we obtain
\begin{eqnarray}
\left\langle {{E^{foc}} \left( {x,y} \right)} \right\rangle && = \frac{\kappa }{{{\lambda ^2}{z^2}}}\sum\limits_{n = 1}^N {{E^{in}}\left( {\xi _n^c,\eta _n^c} \right)  } \nonumber\\
&& \times \iint_{S_n^{b}} {I_n^b\exp \left[ { - \frac{{i2\pi }}{{\lambda z}}\left( {\xi \Delta x + \eta \Delta y} \right)} \right]{\rm{d}}\xi {\rm{d}}\eta },
\end{eqnarray}
where $I_n^b$ is defined as a virtual intensity profile, as a response to $E_n^{seg}$, on the back surface of the medium, and $S_n^b$ represents the area over which $I_n^b$ distributes.

As transverse spreading occurs when the beam propagates through the medium, the transmission beam becomes wider than the incident beam (i.e. $S_n^b$ is larger than $S_n^{in}$). However, when the extent of transverse spreading, which is determined by the scattering coefficient and the thickness of the medium, is smaller than or comparable to the size of a segment, the approximation in Eq. (5) is still valid for Eq. (8). Therefore, the integrals over each of the segments can be added up to be an integral over the back surface,
\begin{eqnarray}
&& \left\langle {{E^{foc}}\left( {x,y} \right)} \right\rangle \nonumber\\
&& = \frac{\kappa }{{{\lambda ^2}{z^2}}} \iint_{S^{b}} {{E^{in}}I_t^b\exp \left[ { - \frac{{i2\pi }}{{\lambda z}}\left( {\xi \Delta x + \eta \Delta y} \right)} \right]{\rm{d}}\xi {\rm{d}}\eta } \nonumber\\
&& = \frac{\kappa }{{{\lambda ^2}{z^2}}}{\cal F}\left[ {E^{in}} \right] \otimes {\cal F}\left[ {I_t^b} \right],
\end{eqnarray}
where ${\cal F}\left[\ \right]$ represents the 2-D Fourier transformation, and $\otimes$ donates the convolution operator. $I_t^b = \sum\limits_{n = 1}^N {I_n^b}$ also refers to the transmitted intensity profile of a unit-amplitude flat-top beam. It signifies that the distribution of the focused filed is determined by the Fourier transformations of $E^{in}$ and $I_t^b$.

Then we consider wavefront shaping with phase-only modulation, where Eq. (7) is modified to be
\begin{equation}
\left\langle {{E^{foc}}\left( {x,y} \right)} \right\rangle  = \sum\limits_{n = 1}^N {{E^{in}}\left( {\xi _n^c,\eta _n^c} \right)\left\langle {\frac{{t_n^ * \left( {0,0} \right)}}{{\left| {{t_n}\left( {0,0} \right)} \right|}}{t_n}\left( {x,y} \right)} \right\rangle } .
\end{equation}
Based on the statistical property of the random light field, the focused field is derived (see details in Supplementary material A). It reveals that the focused fields obtained by amplitude-phase and phase-only modulation have similar formulae only with a coefficient difference. Thus, they can be expressed by an unified formula,
\begin{equation}
\left\langle {{E^{foc}}\left( {x,y} \right)} \right\rangle  = \frac{\kappa C_m }{{{\lambda ^2}{z^2}}}{\cal F}\left[ {E^{in}} \right] \otimes {\cal F}\left[ {I_t^b} \right].
\end{equation}
where $C_m=1$ for amplitude-phase modulation, and $C_m={\pi  \mathord{\left/
 {\vphantom {\pi  {\left[ {4\left\langle {\left| {{t_n}} \right|} \right\rangle } \right]}}} \right.
 \kern-\nulldelimiterspace} {\left[ {4\left\langle {\left| {{t_n}} \right|} \right\rangle } \right]}}$ for phase-only modulation. 
 
With Eq. (11), we first discuss a special case -- the focusing of a flat-top beam, which is also the most common case in the field of wavefront shaping. As ${E^{in}} = {E^{f - t}}$ is a constant, it can be taken from the convolution,
\begin{equation}
\left\langle {{E^{foc}}\left( {x,y} \right)} \right\rangle  = \frac{{\kappa C_m {E^{f - t}}}}{{{\lambda ^2}{z^2}}}{\cal F}\left[ {I_t^b} \right].
\end{equation}
That is to say, the focused filed is proportional to the Fourier transformation of $I_t^b$. This result conforms to the conclusion drawn by I. M. Vellekoop et. al. with the continuous field formalism\cite{Vellekoop2010320}. Significantly, the full width at half maximum (FWHM) of the Airy disk gives the diffraction limit of the opaque lens. Then for the general case of focusing a structured light field, the spatial size of ${\cal F}\left[ {E^{in}} \right]$ is usually much larger than the diffraction limit. Thus, we can simplify Eq. (12) by approximating ${\cal F}\left[ {I_t^b} \right] \approx C \cdot \delta \left({x,y}\right)$, where \emph{C} is a constant coefficient. Then the focused field is written as
\begin{equation}
\left\langle {{E^{foc}}\left( {x,y} \right)} \right\rangle  = \frac{{\kappa C_m C}}{{{\lambda ^2}{z^2}}} {\cal F}\left[ {E^{in}} \right].
\end{equation}

Now we have drawn a significant conclusion from our theory -- the opaque lens Fourier transforms the incident light field during the focusing process, which is like a conventional lens. Therefore, we name this theory as opaque-lens field transformation theory. This theory describes the generalized focusing property of opaque lens. Moreover, it enables the generation of a focus with desired amplitude and phase structure by injecting its inverse Fourier transformation to the opaque lens.

\subsection{\label{sec1_2}The focusing of LG beams}

With the opaque-lens field transformation theory, we can consider the deep transmission of optical vortices through the scattering medium. Laguerre-Gaussian beams (LG beams) are typical realizations of optical vortices, and the complex amplitude of an LG beam is expressed as
\begin{eqnarray}
{\rm{LG}}_{pl}\left( {r,\phi ,z} \right) && = \frac{{{C_{pl}}}}{{\omega \left( z \right)}}{\left[ {\frac{{\sqrt 2 r}}{{\omega \left( z \right)}}} \right]^{\left| l \right|}}\exp \left[ { - \frac{{{r^2}}}{{{\omega ^2}\left( z \right)}}} \right] \nonumber\\
&& \times L_p^{\left| l \right|}\left[ {\frac{{2{r^2}}}{{{\omega ^2}\left( z \right)}}} \right]\exp \left[ { - \frac{{ik{r^2}}}{{2R\left( z \right)}}} \right] \nonumber\\
&& \times \exp \left( {il\phi } \right)\exp \left[ {i\left( {2p + \left| l \right| + 1} \right)\xi \left( z \right)} \right],
\end{eqnarray}
where \emph{p} and \emph{l} represent the radial and azimuthal index respectively, $\omega \left({z}\right)$ donates the beam radius, \emph{k} donates the wave number, $ R \left({z}\right)$ donates the radius of curvature of the beam wavefront, $\xi \left({z}\right)$ represents the Gouy phase shift, $L_p^{\left| l \right|}$ is the generalized Laguerre polynomial, and $C_{pl}$ is a constant coefficient. The azimuthal index \emph{l} is also known as the topological charge (i.e. the quantum number of OAM), and the helical phase structure is described by the term $\exp \left(il\phi \right)$.

The topological charge of an LG beam remains unchanged in Fourier transformation. Therefore, to generate an LG field at the focus, we should input an LG beam with the same topological charges to the opaque lens. After the wavefront is optimized, the light field illuminating the medium is written as
\begin{equation}
{E^m} = {\rm{LG}}_{pl} \times {\Phi ^{foc}},
\end{equation}
where $\Phi ^{foc}$ represents the optimized phase mask for wavefront shaping.

\section{\label{sec3}Experiment}

\subsection{\label{sec3_1}Experimental setup}

To verify the validity of the theory, we experimentally investigated the deep focusing of vortex beam with different topological charges through the scattering medium. The phase-modulation masks for realizing the light focusing are obtained from the transmission matrix (TM), whose elements gives the transmission coefficient relating the input modes and the output modes. The experimental setup for measuring the TM and focusing the vortex beam is sketched in Fig. 2. A liquid-crystal SLM (Hamamatsu, X13138-04, pixel size: 12.5 $\mu$m $\times$ 12.5 $\mu$m) was employed to perform wavefront shaping, which provided a phase-only modulation to the horizontally polarized light with a resolution of 1280 $\times$ 1024. A 532 nm solid-state continuous-wave laser (CNI Laser, MGL-III-532) worked as the light source, and the maximum output power was 200 mW. The beam was expanded to spatially match the SLM, and the polarization state was maintained to be horizontal. The modulated light from the SLM was reflected by a 50:50 beam splitter, and was projected to the pupil of an objective lens (Obj1, Olympus, MPLN 10 $\times$, NA = 0.25) by a 4-f imaging system (f1 = 150 mm, f2 = 75 mm). Then the beam was focused by Obj1 and illuminated the scattering medium which was ground glass (Edmund, 220 grits). The transmission light was imaged onto the complementary metal oxide semiconductor (CMOS) camera (AVT, Mako G-131B, pixel size: 5.3 $\mu$m $\times$ 5.3 $\mu$m) by another objective lens (Obj2, Olympus, MPLN 10 $\times$, NA = 0.25) and an imaging lens (f3 = 180 mm).

\begin{figure}
\includegraphics[width=8.5cm]{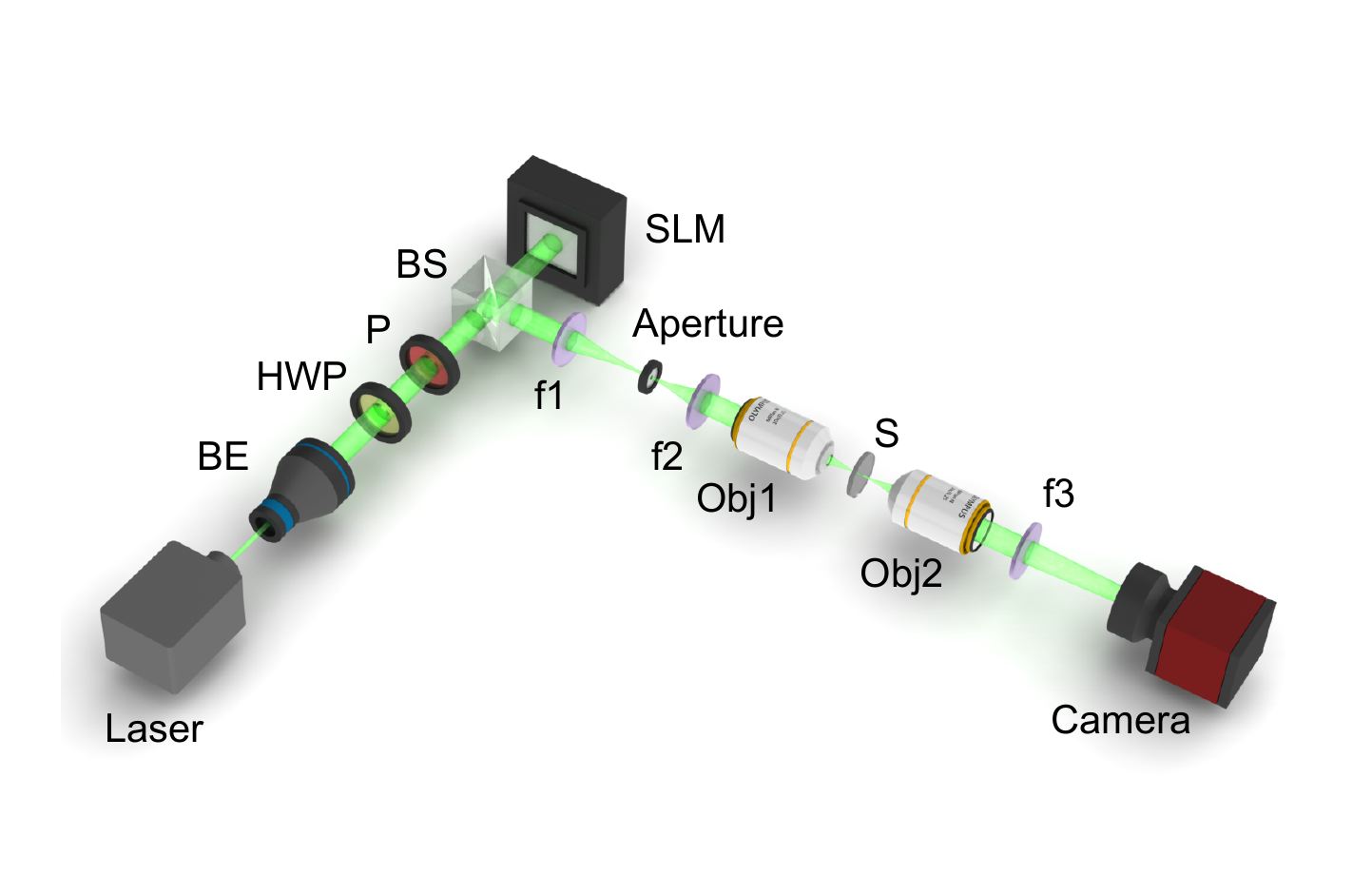}
\caption{\label{fig:2} Schematic diagram of experimental setup for measuring the TM and focusing the vortex beam through the scattering medium. The incident light propagated through the scattering medium after being modulated by the SLM, and the transmission light was measured by the CMOS camera. SLM, spatial light modulator; BE, beam expander; HWP, half-wave plate; P, polarizer; BS, beam splitter; S, scattering medium; f, focusing lens; Obj, objective lens.}
\end{figure}

In theoretical analysis, we have assumed that the SLM is located at the front surface of the scattering medium. In the experimental setup, however, it is necessary to insert the objective lens Obj1 into the middle of the SLM's image plane and the medium. Of course, an available way is to employ a lens to comprise a 4-f system with Obj1. But here we adopted a more compact configuration. We placed the scattering medium 5 mm away from the front edge of Obj1 instead of in the focal region (the front focal length of the objective lens is 10.6 mm). Thus, the light propagation through Obj1 is equivalent to that in a finite distance, where the light field variation cased by the diffraction effect is relatively small. The experimental results in the following section will show that the influence of this variation is negligible. On the other hand, the distance between the focal plane of Obj2 (also acted as the observation plane of the system) and the back surface of the medium was set to be 7 mm.

The TM was measured by sending a set of complete orthogonal fields to the scattering medium and measuring the responses of a series of output modes. We chose 4096 order Hadamard bases as the input fields, and each pixel of the CMOS camera corresponded to an output mode. For each basis the amplitudes and phases of 1024 output modes were recorded by the four-step phase shifting holography, where a static reference light was demanded. We applied the co-propagation scheme proposed in Ref.~\onlinecite{Popoff2010}, where the light illuminating the SLM was divided into signal part and reference part. The signal part was modulated by the central 512 $\times$ 512 pixels of the SLM (subdivided as 64 $\times$ 64 independent control segments) to generate Hadamard bases, while the reference part remained constant. Moreover, to filter out the light noise after modulation (for example, the light reflected by the protective casing), we also loaded a ramp phase pattern on the SLM, and the aperture in the 4-f system only allow the +1st order diffracted light pass. This pattern was only loaded on the signal part during focusing.

\subsection{\label{sec3_2}Experimental results}

The effectivity of the TM was verified by the tight focusing of the flat-top beam. The light beam without phase modulation suffers strong scattering when propagating through the medium, which results in the formation of a diffused wavefront. The partial speckle pattern on the observation plane is depicted in Fig. 3(a). The optimized phase mask $\Phi^{foc}$ for single-spot focusing is shown in Fig. 3(c), which was obtained from the phase-conjugation TM by reshaping the phase of a single row to a 2-D map. When the wavefront was modulated as the mask $\Phi^{foc}$, a sharp focus was generated and shown in Fig. 3(b). In general, the quality of focusing is estimated by peak to background ratio (PBR), which is defined as the ratio of the intensity of the focus to the mean intensity of the background, and the value of PBR in our case was calculated as 1074. In addition, the FWHM of the focus was measured to be 3.18 $\mu$m.

\begin{figure}
\includegraphics[width=8.5cm]{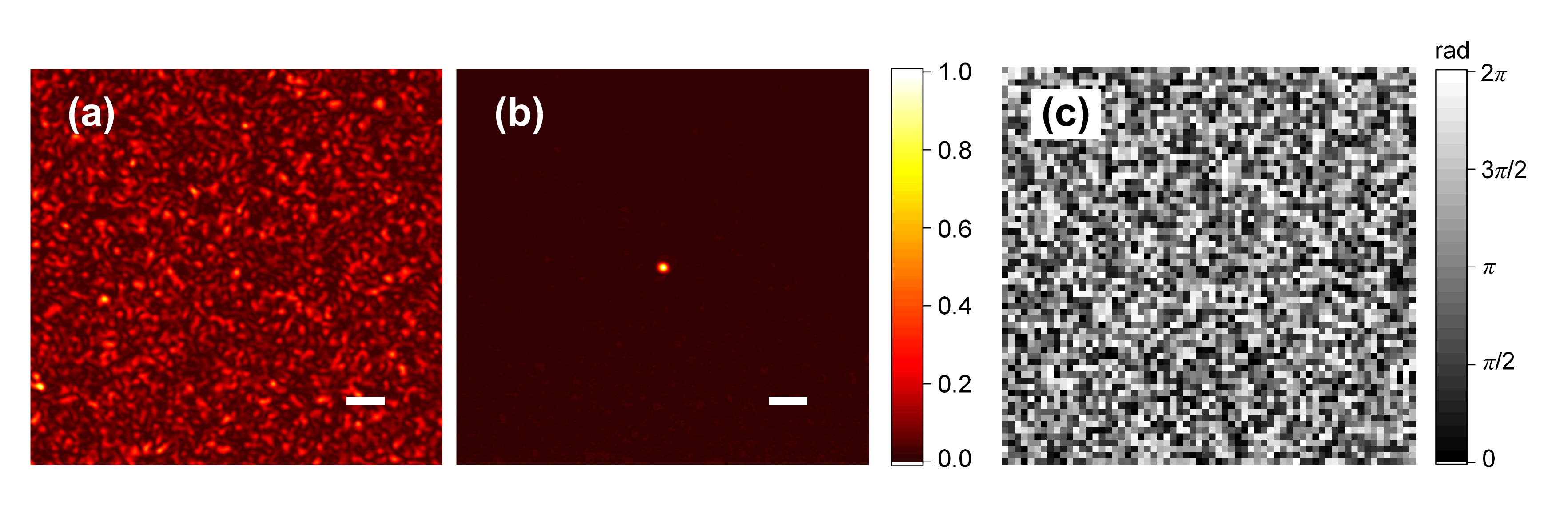}
\caption{\label{fig:3} The generation of a sharp focus by wavefront shaping. (a) The speckle pattern resulting from light scattering. (b) The focus generated by wavefront shaping. (c) The optimized phase mask for single-spot focusing. Scale bar: 10 $\mu$m.}
\end{figure}

\begin{figure*}
\includegraphics[width=15cm]{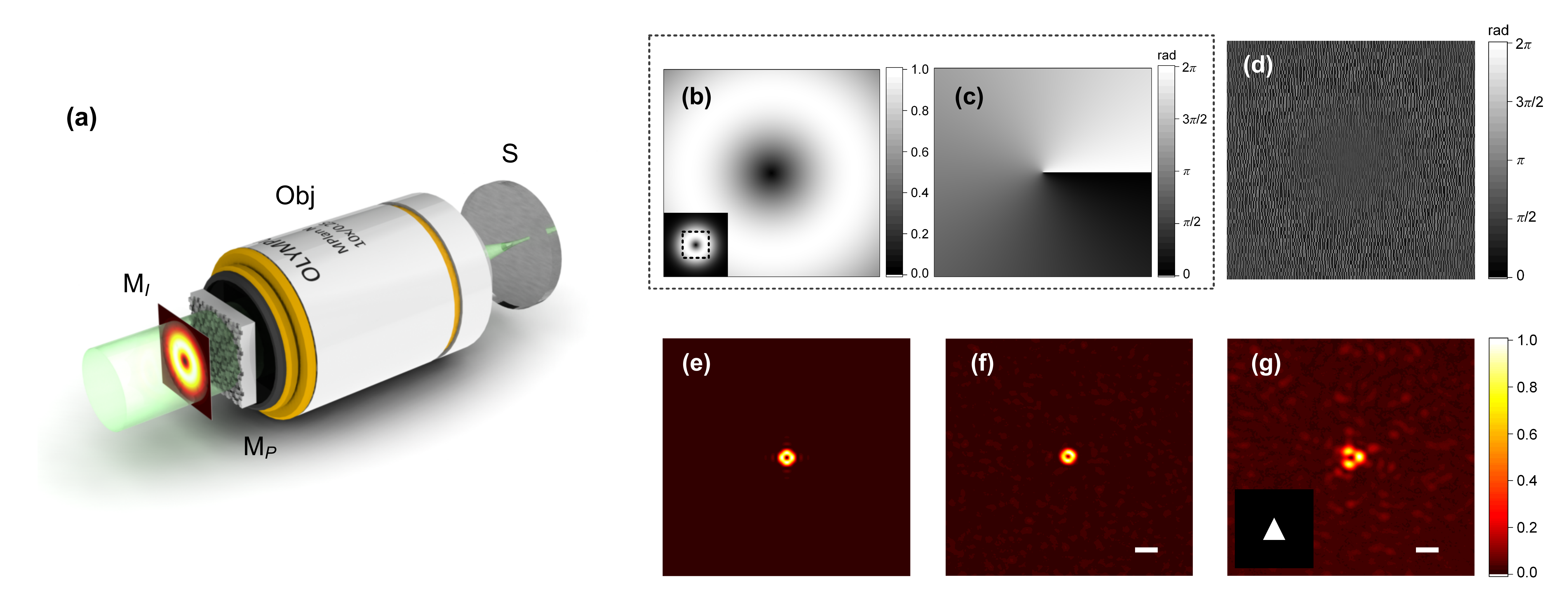}
\caption{\label{fig:4}Experimental results of focusing $\rm{LG}_{01}$ beam through the scattering medium. (a) A simplified schematic diagram of the experiment. The 4-f system that images the SLM to the pupil of the objective lens is omitted. (b)-(c) The amplitude and phase profiles of the $\rm{LG}_{01}$ field inside the effective aperture of the opaque lens. Inset in (b): A complete amplitude profile of the $\rm{LG}_{01}$ field, and the effective aperture is marked by the dashed box. (d) The phase mask for generating the optimized $\rm{LG}_{01}$ field. (e) The 2-D Fourier transformation of the input $\rm{LG}_{01}$ field. (f) The doughnut-shaped focus obtained in the experiment. (g) The diffraction pattern with a triangle aperture. Inset in (g): A diagram of the triangle aperture. Scale bar: 10 $\mu$m.}
\end{figure*}

With the knowledge of the TM, we demonstrate the focusing of the $\rm{LG}_{01}$ beam through the scattering medium. Fig. 4(a) depicts a simplified schematic diagram, where ${\rm{M}}_I$ represents the intensity profile of the $\rm{LG}_{01}$ beam, and ${\rm{M}}_P$ represents the image of the SLM displaying the mask $\Phi^{foc}$. The $\rm{LG}_{01}$ beam entered the objective lens after wavefront shaping. Thanks to the development of the digital modulation technology, both the generation of the $\rm{LG}_{01}$ beam and the wavefront shaping can be simultaneously realized by the same SLM. In other words, the optimized $\rm{LG}_{01}$ field, as described by Eq. (15), can be directly generated by digital modulation. The phase mask for this modulation is presented in Fig. 4(d), and the details of encoding the amplitude and phase by phase-only modulation is shown in Supplementary material B. The $\rm{LG}_{01}$ field on the SLM was designed to be located at the beam waist, and the beam radius $\omega$ was set as 3.5 mm. It is necessary to note that the field was cut off by the effective aperture of the opaque lens, which was decided by the size of the optimized phase mask $\Phi^{foc}$. The amplitude and phase profiles inside the effective aperture are illustrated in Fig. 4(b) and 4(c), respectively. The inset in Fig. 4(b) shows the complete amplitude profile of the $\rm{LG}_{01}$ field, where the effective aperture is marked by the dashed box. With the optimized $\rm{LG}_{01}$ field injected to the medium, a doughnut-shaped focus shown in Fig. 4(f) was generated. Meanwhile, we calculated the 2-D Fourier transformation of the $\rm{LG}_{01}$ field inside the effective aperture, and the intensity profile is shown in Fig. 4(e). It is obvious that the intensity profile of the focus matches well with the theoretical predication. To identify the topological charge of the focus, we inserted a triangular aperture between the lens f3 and the CMOS camera and then recorded the diffraction pattern\cite{Hickmann2010,DeAraujo2011787}. As is shown in Fig. 4(g), a triangular lattice was obtained. There were two spots on each side, which indicates that the focused field is an optical vortex with the topological charge of one. Additionally, we also proposed another method to intuitively exhibit the helical phase of the focus, which we called “self-interference method”. The details of the method and the experimental results are described in Supplementary material C. Both the results of the diffraction and interference methods proved that our approach is effective to transmit the optical vortices through the scattering medium.

Our technology is also applicable to the deep focusing of vortex beam with higher topological charge, and the experimental results for the $\rm{LG}_{02}$ beam and $\rm{LG}_{03}$ beam are presented in Fig. 5. The beam radiuses $\omega$ of both the $\rm{LG}_{02}$ and $\rm{LG}_{03}$ beam were set to be 3.5 mm, and the amplitude and phase profiles inside the effective aperture are respectively illustrated in Fig. 5(a), 5(b), 5(g) and 5(h). With the increase of the topological charge, the LG field distributes in a wider area. Therefore, the intensity distortion in the Fourier domain, which is caused by the diffraction of the effective aperture, gets to be severer. As shown in Fig. 5(d) and 5(j), the intensity profiles in the Fourier domain present as incomplete annuluses, rather than the ideal doughnut-shaped patterns. The phase masks for digitally generating the optimized LG fields are presented in Fig. 5(c) and 5(i), and the resulting foci are shown in Fig. 5(e) and 5(k). Both for the $\rm{LG}_{02}$ and $\rm{LG}_{03}$ beams, the intensity profiles of the foci agree well with the theoretical predications given by Fourier transformation. In particular, the distortions resulting from the aperture diffraction could be observed. As well as the experiment with $\rm{LG}_{01}$ beam, a triangular aperture was used to examine the topological charge of the focused field. The diffraction patterns displayed in Fig. 5(f) and 5(l) suggest that the focused fields have the same topological charges as the input LG beams.

\begin{figure}
\includegraphics[width=8.5cm]{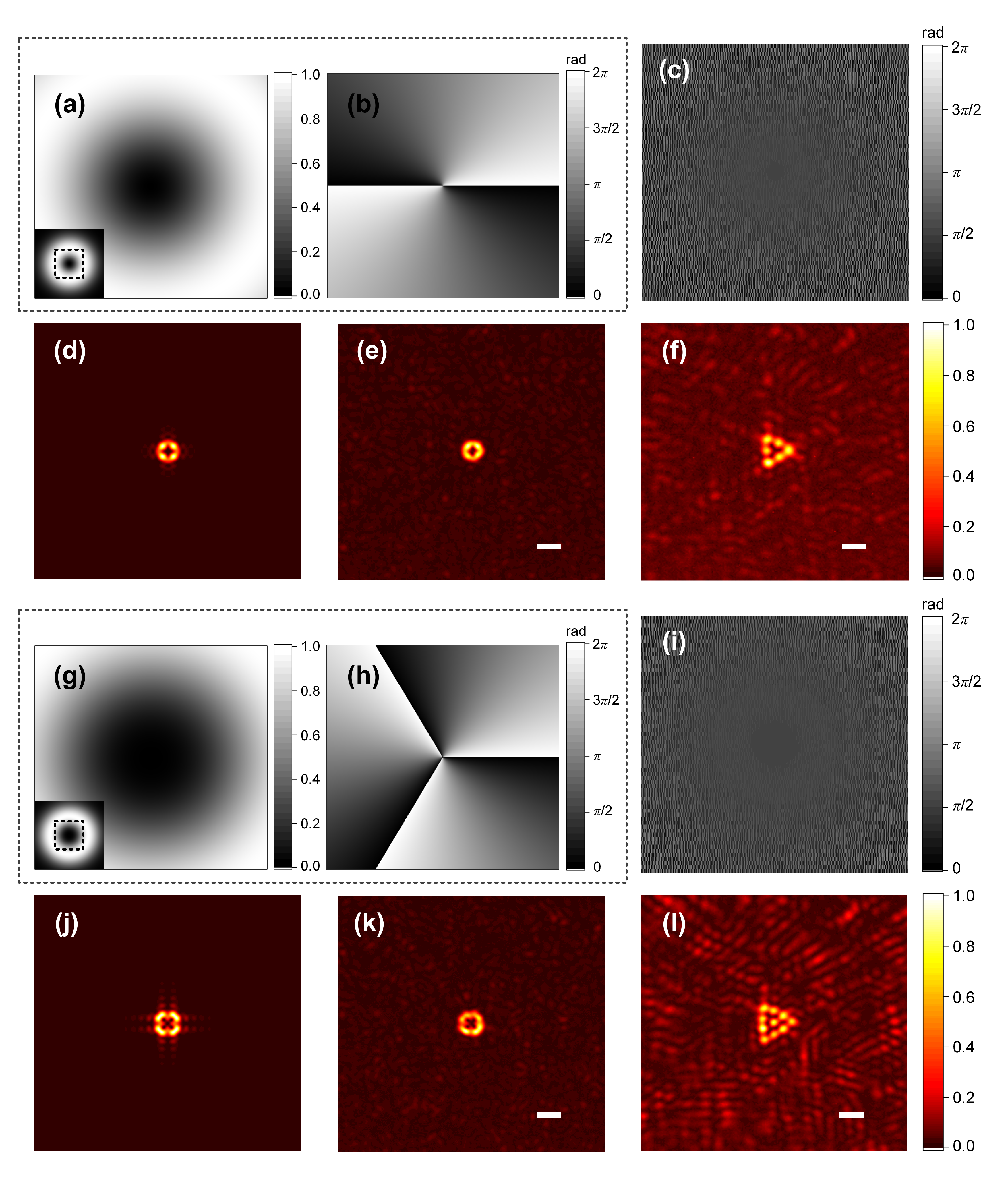}
\caption{\label{fig:5} Experimental results of focusing $\rm{LG}_{02}$ beam and $\rm{LG}_{03}$ beam through the scattering medium. (a)-(b) The amplitude and phase profiles of the $\rm{LG}_{02}$ field inside the effective aperture. (c) The phase mask for generating the optimized $\rm{LG}_{02}$ field. (d) The 2-D Fourier transformation of the input $\rm{LG}_{02}$ field. (e) The focus of $\rm{LG}_{02}$ beam. (f) The triangle-aperture diffraction pattern of the focused $\rm{LG}_{02}$ beam. (g)-(h) The amplitude and phase profiles of the $\rm{LG}_{03}$ field inside the effective aperture. (i) The phase mask for generating the optimized $\rm{LG}_{03}$ field. (j) The 2-D Fourier transformation of the input $\rm{LG}_{03}$ field. (k) The focus of $\rm{LG}_{03}$ beam. (l) The triangle-aperture diffraction pattern of the focused $\rm{LG}_{03}$ beam. Inset in (a) and (g): The complete amplitude profiles of the $\rm{LG}_{02}$ field and $\rm{LG}_{03}$ field, and the effective aperture is marked by the dashed box. Scale bar: 10 $\mu$m.}
\end{figure}

Furthermore, The TM enables multiple-spot focusing through the scattering medium, and the phase mask for multiple-spot focusing is obtained from the linear combination of the masks for focusing to different output modes. Here, we demonstrate the ability to generate the arrays of optical vortices by multiple-spot focusing. With digital modulation, we generated a field written as
\begin{equation}
{E_m} = {\rm{LG}}_{pl} \cdot \sum\limits_{{n_i}} {\Phi _{{n_i}}^{foc}} ,
\end{equation}
where $n_i$ donates the serial number of the output mode. Different arrays of the optical vortices could be obtained by adjusting the combination of output modes. As examples, the horizontal array of three foci with the topological charge of 1 is displayed in Fig. 6(a), while the vertical one is displayed in Fig. 6(b). Besides, the multiple-spot focusing of $\rm{LG}_{02}$ and $\rm{LG}_{03}$ beams was also performed, and the results are presented in Fig. 6(c)-6(f).

\begin{figure}
\includegraphics[width=8.5cm]{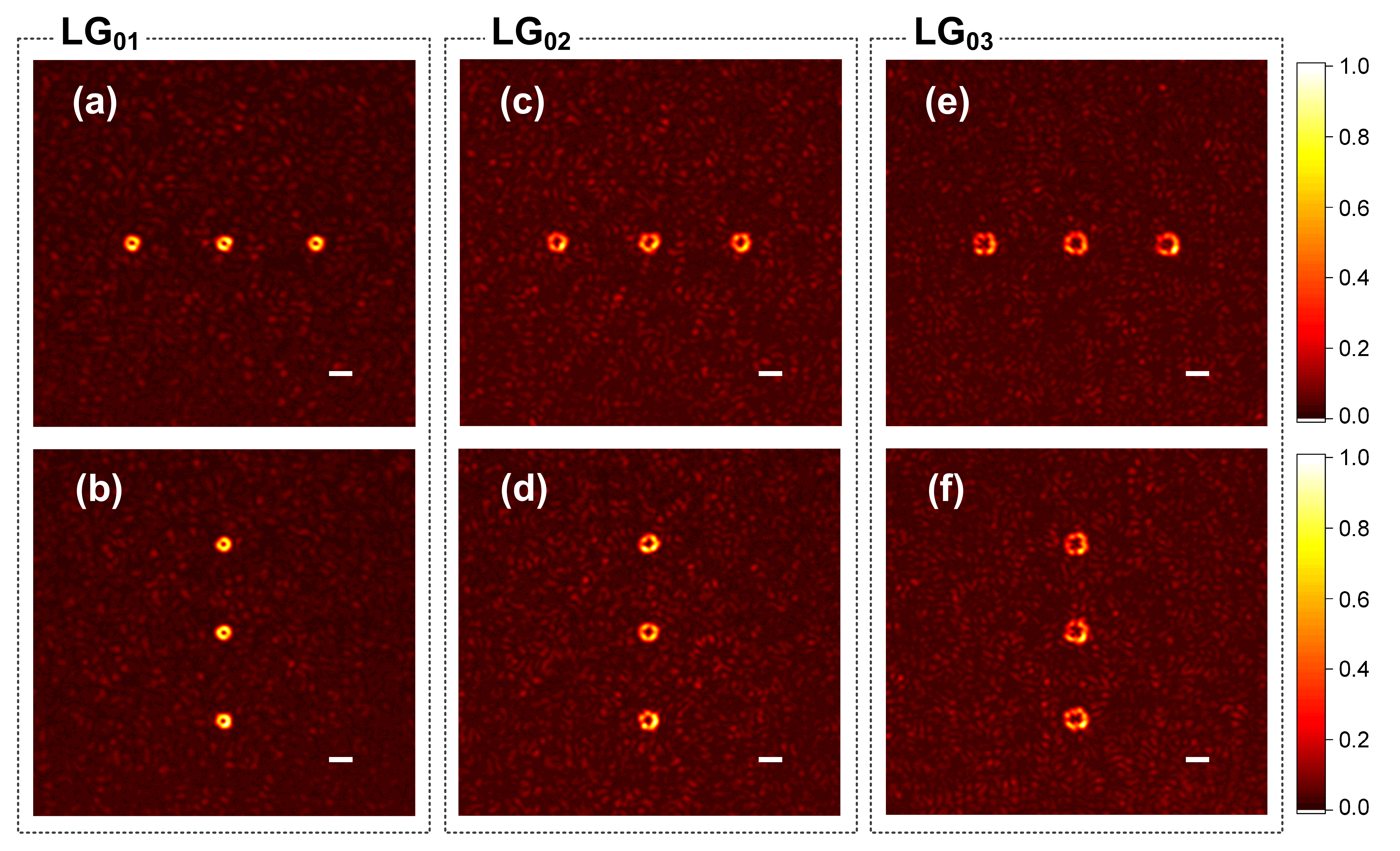}
\caption{\label{fig:6} The generation of multiple optical vortices on the focal plane. (a)-(b) The multiple-spot focusing of $\rm{LG}_{01}$ beam. (c)-(d) The multiple-spot focusing of $\rm{LG}_{02}$ beam. (e)-(f) The multiple-spot focusing of $\rm{LG}_{03}$ beam. Scale bar: 10 $\mu$m.}
\end{figure}

\section{\label{sec4}Discussion}

We have presented, by theory and experiment, that the opaque lens behaves just like a conventional lens when it focuses a structured light field. However, it is necessary to discuss the differences between the opaque lens and the conventional lens. We measured the axial distribution of the focused beam near the focus by moving the objective lens Obj2, and the results are shown in Fig. 7. Although an intended LG field was obtained at the focus of the opaque lens, the light that left the focus presented as a diffused beam rather than an ideal LG beam. This phenomenon is a result of the imperfect wavefront shaping, as the freedom degree of the control in our experiment was far less than that required for the perfect shaping. It is worth mentioning that although this kind of “focused diffused beam” seems to be random, it has the similar diffraction properties to the ideal LG beam. An evidence is given by the triangle-aperture-diffraction experiment demonstrated above. It is inferred that a strong correlation exists between an arbitrary cross-section of the beam and the focus, and any aperture applied to the beam before the focus would result in a diffraction pattern at the focal plane.

For the applications of our technology, the spatial frequency spectrum of the incident field is limited by the acceptance angular range of the memory effect. The maximum acceptance angle is approximately given by ${\theta _{ME}} \approx {\lambda  \mathord{\left/
 {\vphantom {\lambda  {\pi L}}} \right.
 \kern-\nulldelimiterspace} {\pi L}}$, where $\lambda$ represents the wavelength, and \emph{L} represents the thickness of the scattering medium. The spatial spectral components inside the acceptable angular range constitute the focus, while those outside the range form the speckles. The loss of the high-spatial-frequency components not only makes the intended focus field fail to be generated, but reduces the SNR of the system. Therefore, the technology is suitable for focusing the light field dominated by low-spatial-frequency components. On the other hand, for a given spatial bandwidth $\Delta k = {{2\pi \Delta \theta } \mathord{\left/
 {\vphantom {{2\pi \Delta \theta } \lambda }} \right.
 \kern-\nulldelimiterspace} \lambda }$, the thickness must satisfy $L \le {2 \mathord{\left/
 {\vphantom {2 {\Delta k}}} \right.
 \kern-\nulldelimiterspace} {\Delta k}}$ to ensure all the spatial spectral components pass the medium.
 
 Another major limiting factor of our technology is the transverse spreading caused by light scattering. For a Huygens point on the front surface of the medium, the envelope of the ensemble average transmission intensity is given by a PSF\cite{Vellekoop08}. The width of the PSF is defined as the extent of transverse spreading. By introducing the PSF to modify Eq. (9), we could obtain a more accurate expression,
 \begin{eqnarray}
&& \left\langle {{E^{foc}}\left( {x,y} \right)} \right\rangle  = \frac{\kappa }{{{\lambda ^2}{z^2}}} \nonumber\\
&& \times \iint_{S^{b}} {\left[ {{E^{in}} \otimes f} \right]{I^b}\exp \left[ { - \frac{{i2\pi }}{{\lambda z}}\left( {\xi \Delta x + \eta \Delta y} \right)} \right]{\rm{d}}\xi {\rm{d}}\eta },
\end{eqnarray}
where \emph{f} represents the PSF. In the Fourier domain, the spatial spectrum of the incident field is cut off by the transfer function (Fourier transformation of the PSF). The spatial spectral components outside the transfer function are blocked. In order to preserve all the amplitude and phase information, the width of the transfer function must be larger than the spatial bandwidth of the incident field. Assuming that the spatial bandwidth of the incident field is $\Delta k$, then the acceptable maximum extent of transverse spreading is approximately given by ${{W_{\max }^{TS} \propto 1} \mathord{\left/
 {\vphantom {{W_{\max }^{TS} \propto 1} {\Delta k}}} \right.
 \kern-\nulldelimiterspace} {\Delta k}}$. According to the diffusion theory, the extent of transverse spreading is comparable to the thickness of the medium\cite{Vellekoop08}. Hence, the acceptable maximum thickness is written as ${{{L_{\max }} \propto 1} \mathord{\left/
 {\vphantom {{{L_{\max }} \propto 1} {\Delta k}}} \right.
 \kern-\nulldelimiterspace} {\Delta k}}$. This limitation is similar to that posed by the memory effect, because there exists an inherent connection between the transverse spreading and the memory effect\cite{Yllmaz2019}. Finally, our technology will be attractive to the practical applications that transmit complex light fields through thin scattering layers such as eggshell, scarfskin and thin skull.
 
\begin{figure}
\includegraphics[width=8.5cm]{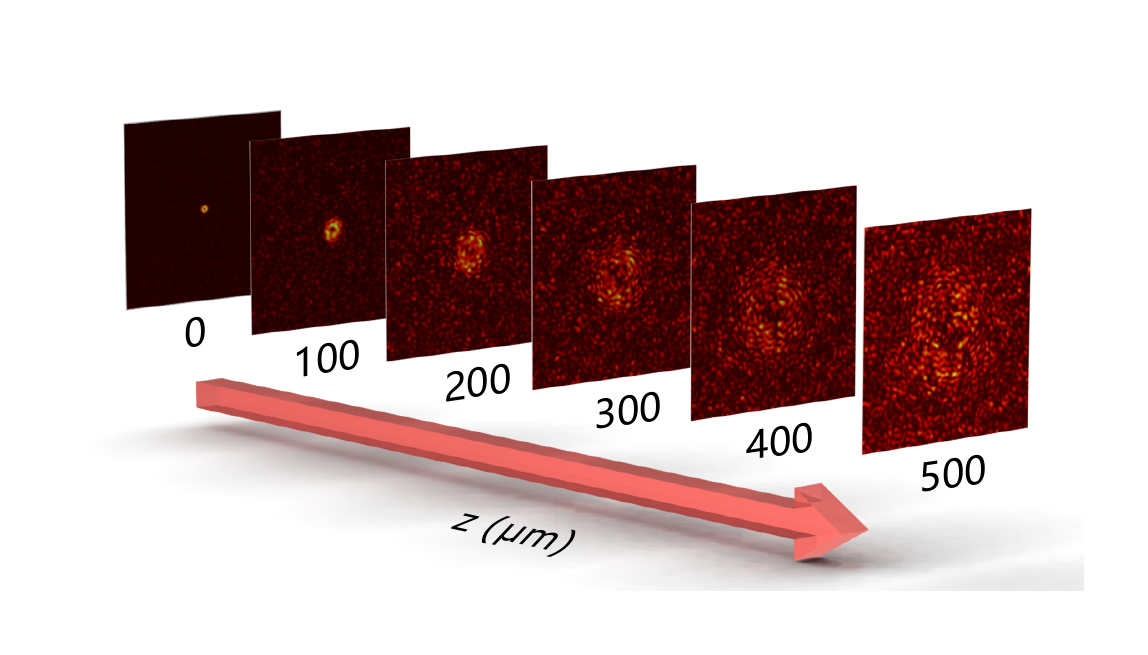}
\caption{\label{fig:7} The axial distribution of the focused beam near the focus. The images displayed here are captured at a space interval of 100 $\mu$m.}
\end{figure}

\section{\label{sec5}Conclusion}

In conclusion, we have proposed a novel theory called “opaque-lens field transformation theory”, which describes the correlation between the focus of the opaque lens and the scalar incident light field. With several approximations, a concise and significant conclusion is drawn -- the opaque lens Fourier transforms the incident filed during focusing process. It provides a theoretical basis for the generation of foci with specified amplitude and phase structures behind scattering media. Guided by this theory, the deep focusing of LG beams with various topological charges were experimentally demonstrated. The doughnut-shaped foci agreed well with the theoretical predications, and the topological charges of the foci were identified with triangle-aperture-diffraction method, verifying the effectivity of this theory in transmitting optical vortices through scattering media. Furthermore, we discussed the two major limiting factors of this technology, i.e. the acceptance angular range of the memory effect and the transverse spreading. They lead to a trade-off between the thickness of the medium and the spatial bandwidth of the incident field. Our approach opens up a new way for the applications of vortex beam in scattering environment. It is of particular interest to the potential bio-photonics technologies such as high-flexibility optical manipulation in biological tissues.

\section*{Supplementary Material}

See Supplementary material for supporting content.

\begin{acknowledgments}
This work was supported by the National Natural Science Foundation of China (NSFC) (Grant Nos. 61905128 and 61875100).
\end{acknowledgments}

\nocite{*}
\providecommand{\noopsort}[1]{}\providecommand{\singleletter}[1]{#1}%
%


\end{document}